\documentclass[prl,aps,twocolumn,superscriptaddress,showpacs]{revtex4}

\usepackage{graphicx}

\begin{document}

\title{Noiseless loss suppression in quantum optical communication}

\author{M. Mi{\v{c}}uda}
\affiliation{Department of Optics, Palack\'{y} University, 17. listopadu 1192/12, CZ-771 46 Olomouc, Czech Republic}

\author{I. Straka}
\affiliation{Department of Optics, Palack\'{y} University, 17. listopadu 1192/12, CZ-771 46 Olomouc, Czech Republic}

\author{M. Mikov\'{a}}
\affiliation{Department of Optics, Palack\'{y} University, 17. listopadu 1192/12, CZ-771 46 Olomouc, Czech Republic}

\author{M. Du{\v{s}}ek}
\affiliation{Department of Optics, Palack\'{y} University, 17. listopadu 1192/12, CZ-771 46 Olomouc, Czech Republic}

\author{N.J. Cerf}
\affiliation{Quantum Information and Communication, Ecole Polytechnique de Bruxelles, CP 165/59, Universit\'e Libre de Bruxelles, 1050 Brussels, Belgium}

\author{J. Fiur\'{a}\v{s}ek}
\affiliation{Department of Optics, Palack\'{y} University, 17. listopadu 1192/12, CZ-771 46 Olomouc, Czech Republic}

\author{M. Je\v{z}ek}
\affiliation{Department of Optics, Palack\'{y} University, 17. listopadu 1192/12, CZ-771 46 Olomouc, Czech Republic}

\begin{abstract}
We propose a protocol for conditional suppression of losses in direct quantum state transmission over a lossy quantum channel.
The method works by noiselessly attenuating the input state prior to transmission through a lossy channel followed by noiseless amplification of the output state.
The procedure does not add any noise  hence it keeps quantum coherence. We experimentally demonstrate it in the subspace spanned by vacuum and single-photon states, 
and consider its general applicability.

\end{abstract}


\maketitle

Quantum communication holds the promise of unconditionally secure information transmission \cite{Scarani09}. However, the distance over which 
quantum states of light can be distributed without significant disturbance is limited due to unavoidable
losses and noise in optical links. Losses, as well as errors or decoherence, may in principle be overcome 
by the sophisticated techniques of quantum error correction \cite{Chiaverini04,Aoki09,Lassen10}, 
entanglement distillation \cite{Pan03,Dong08,Hage08},  and quantum repeaters \cite{Briegel98,Duan01}. 
However, these techniques typically require encoding information into complex multimode entangled states,
processing many copies of an entangled state, and -- even more challenging -- using quantum memories \cite{Lvovsky09,Hammerer10}. 
In stark contrast with the situation for classical communication,  losses in quantum communication
cannot be compensated by amplifying the signal, because the laws of quantum mechanics imply that any deterministic phase-insensitive signal amplification 
is unavoidably accompanied by the addition of noise \cite{Caves82}.

Very recently, however, the concept of heralded noiseless amplification of light \cite{Ralph08} 
was proposed as a way out, relaxing the deterministic requirement. The noiseless amplification 
is formally described by a quantum filter $g^{n}$, where $n$ is the photon number operator and $g>1$ denotes the amplification gain.
The noiseless amplifier thus modulates amplitudes of Fock states $|n\rangle$ by factor $g^n$. This filtration can conditionally increase 
amplitude of a coherent state $|\alpha\rangle$ without adding any noise, 
$g^n|\alpha\rangle \propto |g\alpha\rangle$. Although this cannot be done perfectly because $g^n$ is unbounded,
 faithful noiseless amplification is possible in any finite subspace spanned by the Fock states $|n\rangle$ with $n \leq N$, 
 albeit with a correspondingly low probability scaling as $g^{-2N}$ in the worst case of input vacuum state. 
 With current technology, it has been proven possible to faithfully noiselessly amplify weak coherent states containing 
 mostly vacuum and single-photon contributions \cite{Xiang10,Ferreyrol10,Usuga10,Zavatta11}.

\begin{figure}[!b!]
\centerline{\includegraphics[width=\linewidth]{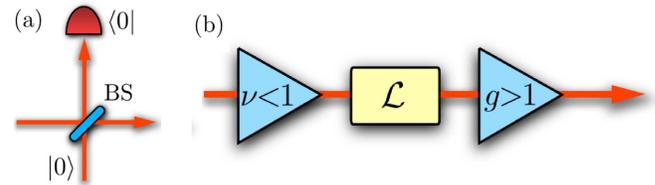}}
\caption{(a)  Implementation of noiseless attenuation with a beam splitter (BS) of amplitude transmittance $\nu$ and a single-photon detector, conditioning on 
projecting onto vacuum $|0\rangle$. (b) Conditional noiseless loss suppression in direct state transmission over a lossy channel $\mathcal{L}$ 
by a combination of noiseless attenuation with transmitance $\nu$ and noiseless amplification with gain $g$. }
\end{figure}

The noiseless amplifier can improve the performance of quantum key distribution protocols \cite{Gisin10,Blandino12,Fiurasek12,Walk12} and 
it can also be used to distribute high-quality entanglement over a lossy channel \cite{Ralph08,Ralph11}.
Beyond that, the noiseless amplifier is not useful to suppress losses in direct transmission of arbitrary quantum states 
because it is not the inverse map of a lossy channel $\mathcal{L}$. As a matter of fact, any superposition of Fock states that is not 
a coherent state is mapped by $\mathcal{L}$ onto a mixed state, and this added noise cannot be eliminated
by noiseless amplification. 

Here, we find a solution to this fundamental problem by introducing the concept of noiseless attenuation, which can be viewed
as a heralded but reversible type of loss in the sense that the state becomes closer to vacuum, while its purity and quantum coherence are preserved. 
Mathematically, noiseless attenuator is described by an operator $\nu^n$ with $\nu<1$.
This filtering can be accomplished with the help of a beam splitter with amplitude transmittance $\nu <1$ and a single-photon detector monitoring 
the auxiliary output port of the beam splitter, see Fig. 1(a). If the detector does not register any photon, then the amplitudes of Fock states $|n\rangle$
are attenuated according to $|n\rangle\rightarrow \nu^n|n\rangle$.
The noiseless attenuator transforms $|\alpha\rangle \rightarrow |\nu\alpha\rangle$,  but, unlike $\mathcal{L}$, it is the proper inverse map of the noiseless amplifier 
$g^n$ with $g=1/\nu$.

In this Letter, we prove that a suitable combination of noiseless attenuation and amplification provides a powerful tool 
to conditionally suppress losses in channel $\mathcal{L}$ to an arbitrary extent without adding noise. 
Our scheme works as shown in Fig. 1(b). Before transmission through $\mathcal{L}$, the input state is noiselessly attenuated 
with transmittance $\nu$. Intriguingly, this input-state preprocessing has the effect of preferentially reducing the weight of the
Fock states that have a higher chance of being affected afterwards by losses in $\mathcal{L}$.
After transmission through  $\mathcal{L}$, the state is noiselessly amplified with gain $g=1/(\nu\tau)$,
where $\tau$ is the amplitude transmittance of $\mathcal{L}$. 
In the limit $\nu \rightarrow 0$, this procedure conditionally converts the lossy channel $\mathcal{L}$
into a perfect lossless channel on the subspace where noiseless amplification $|n\rangle\rightarrow g^n|n\rangle$ 
is faithfully performed.

In order to provide more insight into our protocol, 
let us consider the simple, yet important case of an input state formed by a superposition of vacuum and single-photon states, 
$|\psi\rangle=c_0|0\rangle+c_1|1\rangle$. At the output of $\mathcal{L}$, we get the mixed state
\begin{equation}
\rho_{\mathrm{loss}}=|\tilde{\psi}\rangle\langle\tilde{\psi}|+(1-\tau^2)|c_1|^2|0\rangle\langle 0|,
\end{equation}
where $|\tilde{\psi}\rangle=c_0|0\rangle+\tau c_1|1\rangle$.  A naive compensation of losses by noiseless amplification of 
$\rho_{\mathrm{loss}}$ with gain $g=1/\tau$ results in the transformation $|0\rangle\rightarrow |0\rangle$, $|1\rangle\rightarrow g|1\rangle$, and 
yields the state 
\begin{equation}
\rho_{\mathrm{amp}} \propto |\psi\rangle\langle\psi|+(1-\tau^2)|c_1|^2|0\rangle\langle 0|.
\end{equation}
Note that there remains an extra vacuum noise term  proportional to $(1-\tau^2)|c_1|^2$. 
This noise term could in principle be further suppressed by amplification with a gain higher than $1/\tau$, but such an approach 
would over-amplify the single-photon contribution.

The right solution is to preprocess the state via noiseless attenuation before the lossy channel $\mathcal{L}$. The
effective input state of $\mathcal{L}$ then becomes $|\psi_{\mathrm{eff}}\rangle=c_0|0\rangle+\nu c_1|1\rangle$,
and the output state after attenuation, transmission, and amplification with $g=1/(\nu\tau)$ reads
\begin{equation}
\rho_{\mathrm{out}}\propto |\psi\rangle\langle\psi|+(1-\tau^2)\nu^2|c_1|^2|0\rangle\langle 0|.
\end{equation}
This has reduced the unwanted vacuum noise term by a factor of $\nu^2$. In the 
limit $\nu\rightarrow 0$, this term vanishes and the output state becomes equal to the input pure state $|\psi\rangle$.
The procedure thus conditionally converts a lossy channel $\mathcal{L}$ into a channel that is arbitrarily close to the 
identity channel $\mathcal{I}$.

Our protocol formally resembles the scheme for the suppression of qubit decoherence due to zero-temperature
energy relaxation by using partial quantum measurements \cite{Korotkov,Lee}, but, importantly, it compensates losses instead of qubit decoherence
and can be extended to arbitrarily large Hilbert space as we show now. Lossy channel $\mathcal{L}$ with inputs restricted to the subspace spanned by Fock states 
$|n\rangle$ with $n\leq N$  can be described by a finite number $N+1$  of Kraus operators $A_j$,
\begin{equation}
\rho_{\mathrm{out}}=\mathcal{L}(\rho_{\mathrm{in}})=\sum_{j=0}^N A_j\rho_{\mathrm{in}}A_j^\dagger,
\end{equation}
 where $ A_{j}=\sum_{m=0}^{N-j}\sqrt{\frac{(m+j)!}{m! \, j!}}(1-\tau^2)^{j/2} \tau^m |m\rangle\langle m+j|$
accounts for loss of $j$ photons in a channel. 
Assuming that the noiseless amplification is performed perfectly on the subspace of the first $N+1$ Fock states by filter $G_N(g)=g^{-N}\sum_{n=0}^N g^n |n\rangle\langle n|$,
the effective channel $\mathcal{M}$ formed by sequence of noiseless attenuation, losses, and noiseless amplification reads,
\begin{equation}
\mathcal{M}(\rho_{\mathrm{in}})=  \sum_{j=0}^N G_N(g) A_j\nu^{n}\rho_{\mathrm{in}}\nu^{n}A_j^\dagger G_N(g),
\end{equation}
with $g=1/(\nu\tau)$. Due to the structure of Kraus operators we find that $G_N(g) A_j\nu^{n}=g^{-N}\nu^{j} A_0^{-1} A_{j}$ and the effective channel 
can be expressed as,
\begin{equation}
\mathcal{M}(\rho_{\mathrm{in}})= g^{-2N} \rho_{\mathrm{in}}+g^{-2N}\sum_{j=1}^N \nu^{2j}B_j\rho_{\mathrm{in}}B_j^\dagger,
\end{equation}
where $B_{j}=A_{0}^{-1}A_j$, and the inverse $A_0^{-1}=\sum_{n=0}^N \tau^{-n}|n\rangle\langle n|$ exists on the considered finite dimensional subspace.
We can clearly see that the combination of attenuation and noiseless amplification progressively suppresses $j$-photon losses by a factor of $\nu^{2j}$ and in the 
limit $\nu\rightarrow 0$ we approach the identity channel, $\mathcal{M}\rightarrow \mathcal{I}$. Success probability of the protocol is state dependent,
\begin{equation}
P_{\mathrm{succ}}=g^{-2N}+g^{-2N}\sum_{j=1}^N\nu^{2j}\mathrm{Tr}[B_{j}^\dagger B_j \rho_{\mathrm{in}}],
\label{Psuccdef}
\end{equation}
and $P_{\mathrm{succ}}$ is lower bounded by $g^{-2N}$.

\begin{figure}[!b!]
\centerline{\includegraphics[width=\linewidth]{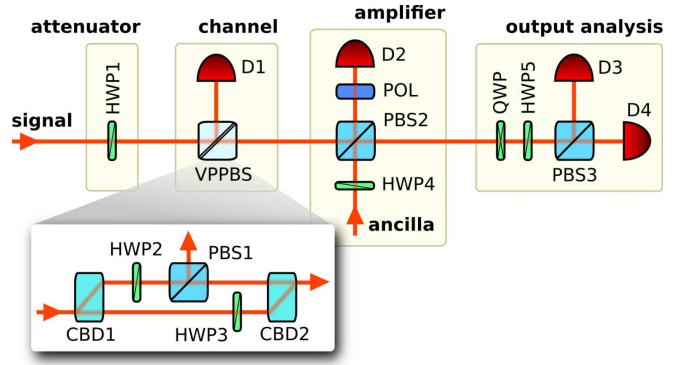}}
\caption{Experimental setup. Correlated signal and idler photons with wavelength of 810 nm are generated in the process 
of spontaneous parametric downconversion in a nonlinear $\beta$-BaB${}_2$O${}_4$ crystal pumped by a laser diode \cite{Jezek11} (not shown) 
and injected into a linear optical setup consisting of polarizing beam splitters (PBS), half-wave plates (HWP), quarter-wave plate (QWP),
polarizer (POL), and a variable partially polarizing beam splitter (VPPBS). Photons are detected with four single-photon
 detectors D${}_j$. The VPPBS is constructed from a pair of calcite beam displacers (CBD) with two HWPs and
a PBS in between. The device spatially separates and subsequently re-combines horizontally and vertically polarized beams.
It introduces tunable losses in the vertical-polarization component by rotation of HWP$_2$.}
\end{figure}

We have experimentally demonstrated this protocol for superpositions of vacuum and single-photon states.
 In the experimental setup, shown in Fig. 2, a correlated photon pair is generated, and
the signal photon serves as a probe of the lossy channel $\mathcal{L}$ while the idler photon 
drives noiseless amplification. We characterize channel $\mathcal{L}$ and the loss suppression mechanism by using the
Choi-Jamolkowski isomorphism \cite{Jamiolkowski72,Choi75} between quantum channels and bipartite states.
We exploit the polarization degree of freedom of the signal photon. 
The vertically (V) polarized mode is transmitted through $\mathcal{L}$, while the horizontally (H) polarized mode is transmitted through 
a reference identity channel $\mathcal{I}$. If the signal photon is initially diagonally polarized, 
 $|\Psi^{+}\rangle=\frac{1}{\sqrt{2}}(|1\rangle_V|0\rangle_H+|0\rangle_V|1\rangle_H)$, then we obtain at the output 
a two-mode state $\chi_{\mathcal{L}}=[ \mathcal{L}_V\otimes \mathcal{I}_H ](\Psi^{+})$ that is isomorphic to the channel $\mathcal{L}$, 
\begin{equation}
\chi_{\mathcal{L}}=\frac{1-\tau^2}{2}|00\rangle\langle 00|+\frac{1}{2}(|01\rangle+\tau|10\rangle)(\langle 01|+\tau\langle 10|),
\label{chiloss}
\end{equation}
where the subscripts $H$ and $V$ were suppressed for simplicity. Although $\chi_{\mathcal{L}}$ is a two-qubit density matrix, its support 
is  restricted to a three-dimensional subspace spanned by $|00\rangle$, $|10\rangle$, and $|01\rangle$. The state $|11\rangle$
is absent because no photons are generated in the passive channel $\mathcal{L}$. This property holds even if the channel is combined with 
noiseless attenuation and amplification because these operations only modify Fock state amplitudes. 
In the experimental tomographic reconstruction of $\chi_{\mathcal{L}}$ we can therefore restrict ourselves 
to the above three-dimensional subspace. The only nonzero off-diagonal elements of $\chi_{\mathcal{L}}$ are $\langle 01|\chi_{\mathcal{L}}|10\rangle$ and its conjugate.
 Since their phase can be set to zero by a suitable phase shift $e^{in\phi}$, we can represent $\mathcal{L}$ by a real $\chi_{\mathcal{L}}$ 
 without any loss of generality.
 
Noiseless amplification is accomplished by two-photon interference on polarizing beam splitter PBS$_2$
that transmits horizontally polarized modes and reflects vertically polarized modes \cite{Xiang10}. 
The state to be amplified is injected into vertically polarized 
 mode of the first input port of PBS$_2$. An idler photon prepared in linearly polarized state 
$\cos\theta|0\rangle_V|1\rangle_H+\sin\theta|1\rangle_V|0\rangle_H$ is injected into the second input port of PBS$_2$.
 Noiseless amplification is successful if a single photon emerges in the 
 auxiliary output port of PBS$_2$ and is projected onto diagonally linearly polarized state $|\Psi^{+}\rangle$ which is heralded by a click of detector $D_2$.
 Amplification gain of this scheme is given by $g=\tan \theta$ and can be tuned by rotating HWP4.
 Note that our implementation of the noiseless amplifier has a success probability lower by a factor of $\frac{g^2}{2(1+g^2)}$ 
 than the optimal filter $G_1=g^{-1}|0\rangle\langle 0|+|1\rangle\langle 1|$.
 Improvement by a factor of $2$ could be obtained by taking into account also projections onto 
 anti-diagonally polarized state and introducing active feed-forward that performs $\pi$ phase shift on 
 vertically polarized signal mode in this case \cite{Mikova12}.
 
Noiseless attenuation is, in this proof-of-principle experiment, simply equivalent to preparing the signal photon in a suitably linearly polarized state 
$\nu^{\,n_V} |\Psi^{+}\rangle\propto\frac{1}{\sqrt{1+\nu^2}}(|0\rangle_V|1\rangle_H+\nu|1\rangle_V|0\rangle_H)$, which is accomplished by half-wave plate HWP$_1$. 
Note that at the cost of increased technological complexity, the component noiseless attenuator and amplifier 
can both be accomplished with high fidelity in a  heralded manner, even with inefficient single-photon detectors, 
using combination of photon addition and subtraction \cite{Marek10,Fiurasek09,Zavatta11}.

\begin{figure}[!t!]
\centerline{\includegraphics[width=\linewidth]{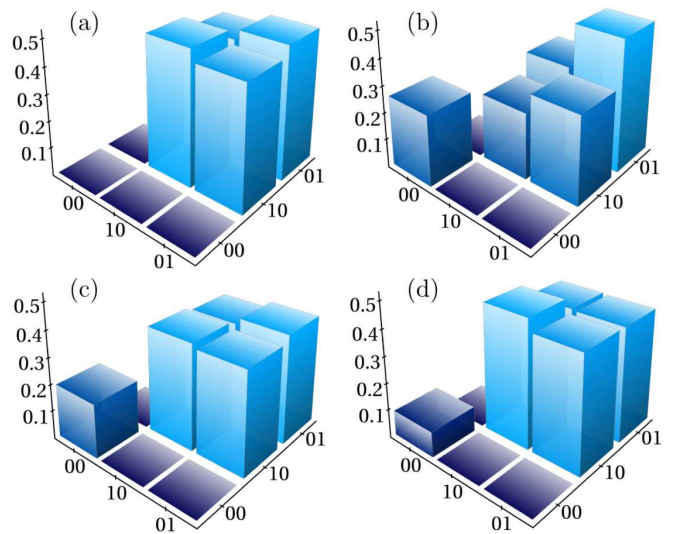}}
\caption{Reconstructed channel matrices. Shown are experimentally determined $\chi_{\mathcal{L}}$ matrices characterizing four different channels 
for input states restricted to the subspace spanned by vacuum and single-photon states. 
(a) Identity channel. (b) Lossy channel with amplitude transmittance $\tau=1/\sqrt{2}$. (c) Lossy channel compensated by noiseless amplification with gain $g=\sqrt{2}$.
(d) Lossy channel compensated by the combination of noiseless attenuation with $\nu=1/\sqrt{2}$ at the input and noiseless amplification 
with gain $g=2$ at the output. All matrices are normalized such that $\mathrm{Tr}[\chi_{\mathcal{L}}]=1$.}
\label{setupfig}
\end{figure}

The state analysis block including detectors $D_3$ and $D_4$ serves for a full tomographic analysis of the polarization state 
of the output signal photons. Moreover, we employ detector $D_1$ to monitor the fraction of photons that are lost in channel $\mathcal{L}$.
Note that, in contrast to noise-reduction schemes based on measurements of the environment \cite{Sciarrino09}, detector $D_1$ 
is not needed for the protocol itself and only serves here for channel characterization. 
We  measure the two-photon coincidences  $D_1 \& D_2$, $D_3\& D_2$, and $D_4\& D_2$ for different settings of the wave-plates 
in the output analysis block, and from these data we completely determine $\chi_{\mathcal{L}}$.

In Fig. 3, we plot the reconstructed  matrices $\chi_{\mathcal{L}}$ for four different channels. 
Panel (a) shows $\chi_{\mathcal{L}}$ when no losses are inserted in the path of vertically-polarized signal photons. 
The reconstructed operator is very close to the identity-channel matrix $\chi_{\mathcal{I}}=|\Psi^{+}\rangle\langle\Psi^{+}|$.
The similarity of a channel $\mathcal{L}$ with respect
to the identity channel $\mathcal{I}$ is measured by the channel fidelity $F=\langle\Psi^{+}|\chi_{\mathcal{L}}|\Psi^{+}\rangle$,
and we obtain here $F=0.958\pm 0.002$.  
Then, the $\chi_{\mathcal{L}}$ matrix of a channel  $\mathcal{L}$ with $50\%$ losses ($\tau=1/\sqrt{2}$) is depicted in Fig. 3(b). 
Losses introduce imbalance between the amplitudes of $|01\rangle$ and $|10\rangle$ states, and give rise to a nonzero probability
for the $|00\rangle$ state, which represents the fraction of lost photons in the channel.
If we attempt to compensate these losses by noiseless amplification with  gain $g=\sqrt{2}$,
we obtain the channel in Fig. 3(c). The amplification balances the amplitudes of $|01\rangle$ and $|10\rangle$ states, but
there remains some population in the $|00\rangle$ state. This unwanted noise can be further suppressed if we include 
noiseless attenuation. The resulting channel for $\nu=1/\sqrt{2}$ and $g=2$ is shown in Fig. 3(d). 
In contrast with Fig. 3(c), the noise term is reduced while quantum coherence is preserved, as witnessed by the off-diagonal terms
in the subspace spanned by $|01\rangle$ and $|10\rangle$.

\begin{figure}
\centerline{\includegraphics[width=\linewidth]{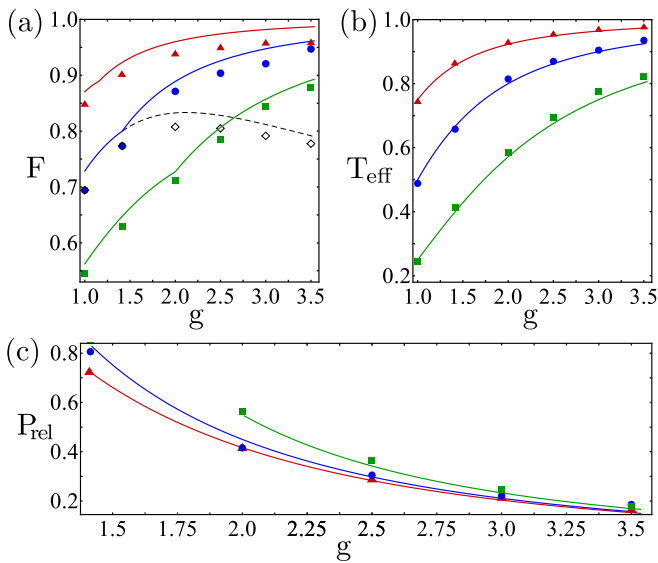}}
\caption{Performance of the noiseless loss suppression scheme.  Fidelity $F$ of the resulting quantum channel (a), effective channel transmittance 
$T_{\mathrm{eff}}$ (b),  and relative success probability $P_{\mathrm{rel}}$ of the protocol (c)
are plotted as functions of the amplification gain $g$ with the attenuation set to $\nu=\min[1/(g\tau),1]$. 
Symbols represent experimental results
for three levels of losses: $\tau^2=75\%$ (red triangles), $\tau^2=50\%$ (blue circles) and $\tau^2=25\%$ (green squares).
Solid lines indicate corresponding theoretical predictions. Diamonds and dashed line in panel (a) provide channel fidelity for the case 
without noiseless attenuation ($\nu=1$) and $50\%$ losses.
Statistical errors are smaller than the symbol size.
The experimental data for $P_{\mathrm{rel}}$ are multiplied by $\frac{1}{2}(1+\nu^2)$ in order to 
compensate for the fact that the noiseless attenuation $\nu$ was included in the input state preparation.}
\label{setup}
\end{figure}

We have systematically investigated the performance of the protocol  as a function of the amplification gain $g$ 
for three different levels of losses: $25\%$, $50\%$, and $75\%$.  
The fidelity $F$ of the resulting quantum channel, plotted in Fig. 4(a), 
monotonically grows with $g$, and theory (solid lines) predicts $F\to 1$ in the high gain limit. The experimentally observed 
$F$ saturates at values slightly below $1$, which can be attributed mainly to imperfect two-photon interference on PBS$_2$. 
The measured visibility of Hong-Ou-Mandel dip \cite{Hong87} $V=0.947\pm 0.002$ is in good agreement with the observed saturation.
For comparison, we also plot data for the naive loss compensation strategy based solely on noiseless amplification without
input state preprocessing ($\nu=1$). The results shown as diamonds demonstrate the fundamental limitation of this strategy. 
With increasing gain, the channel fidelity reaches the maximum $F_{\mathrm{max}}=(3-\tau^2)/(4-2\tau^2)$ for 
$g_{\mathrm{opt}}=(2-\tau^2)/\tau$, and then drops down due to over-amplification of the single-photon part of the state. 
We define an effective channel transmittance  $T_{\mathrm{eff}}$ as the conditional probability that a photon injected into the channel 
emerges at the output. Figure 4(b) demonstrates that $T_{\mathrm{eff}}$ monotonically increases with $g$ 
and approaches unity in the high gain limit. 

The noiseless loss suppression is a conditional operation, therefore its success probability is a crucial parameter. 
Assuming pure input state $c_0|0\rangle+c_1|1\rangle$ we obtain
\begin{equation}
P_{\mathrm{succ}}=|c_0|^2g^{-2}+|c_1|^2\nu^2[\tau^2+(1-\tau^2)g^{-2}]
\label{Psucc01}
\end{equation}
which agrees with Eq. (\ref{Psuccdef}) if $g=1/(\nu \tau)$.
The actual experimental success probability is significantly lower because it is 
reduced by imperfect collection efficiency $\eta_C$ of the idler photon and low overall detection efficiency $\eta_D$ of the heralding detector D$_2$.  
We can only roughly estimate $\eta_C\eta_D\approx 0.1$. On the other hand, by taking ratio of the measured total coincidence rates for a given $\tau$ and $g$
and for the identity channel with $\tau=1$ we can reliably estimate a relative success probability normalized such that 
$P_{\mathrm{rel}}=1$  for the identity channel. We expect this relative success probability to be equal to (\ref{Psucc01}).
Relative success probability for the probe state $|\Psi^{+}\rangle$ is plotted in Fig. 4(c). It is in good agreement with theoretical predictions
obtained by setting $|c_0|^2=|c_1|^2=\frac{1}{2}$ in Eq. (\ref{Psucc01}) and it scales as $g^{-2}$ as expected.

In summary, we have experimentally demonstrated a protocol for the conditional noiseless suppression of losses 
in  quantum optical channels using quantum filters at the input and output of the channel. The procedure is universally applicable 
and can enable faithful transmision of fragile highly non-classical or entangled states of light over lossy channel.
The state transmitted over loss-compensated channel can be made fully available for further processing by employing more experimentally 
demanding heralded version of noiseless attenuation and amplification based on photon addition and subtraction \cite{Marek10,Fiurasek09,Zavatta11}. 
We anticipate numerous potential applications of the present scheme in quantum communication, quantum metrology, 
and other fields where loss reduction is essential for optimal performance.

\acknowledgments
This work was presented at the Continuous variable quantum information processing workshop (CVQIP'11, Paris, Sep. 26-27, 2011).
The work was supported by Projects No. MSM6198959213 and No. LC06007 of the Czech Ministry of Education, 
by Palacky University (PrF-2012-019) and by Czech Science Foundation (202/09/0747). 
N.J.C. acknowledges support from the F.R.S.-FNRS under project HIPERCOM.

\end{document}